\documentclass[conference]{IEEEtran}
\IEEEoverridecommandlockouts
\usepackage{cite}
\usepackage{amsmath,amssymb,amsfonts}
\usepackage{algorithmic}
\usepackage{graphicx}
\usepackage{textcomp}
\usepackage{xcolor}
\def\BibTeX{{\rm B\kern-.05em{\sc i\kern-.025em b}\kern-.08em
    T\kern-.1667em\lower.7ex\hbox{E}\kern-.125emX}}
\begin{document}

\title{Mixed Attention Transformer Enhanced Channel Estimation for Extremely Large-Scale MIMO Systems
\thanks{This work was supported in part by the National Science Foundation of China under Grant 62101253, the Natural Science Foundation
of Jiangsu Province under Grant BK20210283, the Fundamental
Research Funds for the Central Universities, NO. NS2024023, and the open research fund of National Mobile Communications Research Laboratory, Southeast University (No. 2022D08). (Corresponding author: Peihao Dong.)}
}

\author{\IEEEauthorblockN{Shuangshuang Li$^*$, Peihao Dong$^*$$^\dagger$}
	\IEEEauthorblockA{$^*$ College of Electronic and Information Engineering, Nanjing University of Aeronautics and Astronautics, Nanjing 211106, China\\
		$^\dagger$ National Mobile Communications Research Laboratory, Southeast University, Nanjing 211111, China\\
		Email: \{lishsh, phdong\}@nuaa.edu.cn}
}

\maketitle

\begin{abstract}
Extremely large-scale massive multiple-input multiple-output (XL-MIMO) is one of the key technologies for next-generation 
wireless communication systems. However, acquiring the accurate high-dimensional channel matrix of XL-MIMO 
remains a pressing challenge due to the intractable channel property and the high complexity.
In this paper, a Mixed Attention Transformer based Channel Estimation Neural Network (MAT-CENet) is developed, which is inspired
by the Transformer encoder structure as well as organically integrates the feature map attention and spatial attention mechanisms to better grasp the unique characteristics of the XL-MIMO
channel. By incorporating the multi-head attention layer as the core enabler, the insightful feature importance 
is captured and exploited effectively. A comprehensive complexity analysis for the proposed MAT-CENet is also provided. Simulation results show that MAT-CENet
outperforms the state of the art in different propagation scenarios of near-, far- and hybrid-fields.

\end{abstract}


\section{Introduction}
Extremely large-scale massive multiple-input multiple-output (XL-MIMO) has emerged as an innovative technique for the sixth generation 
wireless communication\cite{b1}. By deploying a huge number of antennas at the base station, XL-MIMO significantly enhances the 
communication capacity and reliability. Benefiting from the extensive antennas, XL-MIMO excels in terms of the spectral 
efficiency, energy efficiency, and interference mitigation. In XL-MIMO, the base station needs to acquire the channel state information (CSI) 
of a large number of antennas, resulting in the high-dimensional channel matrices. Compared to traditional 
massive MIMO systems, the channel matrices in XL-MIMO not only have higher dimensions but also exhibit more intractable characteristics. 
When dealing with such a large-scale antenna system, the computational load increases dramatically, requiring efficient algorithms and 
powerful computing resources to perform channel estimation. In XL-MIMO systems, due to the vast array of antennas, near-field 
effect becomes more pronounced, which impacts estimation accuracy.

Traditional channel estimation methods typically rely on linear models and statistical assumptions. For example, a scheme based on the least square (LS) 
method was developed in \cite{b3}, but this method can only achieve a low received signal-to-noise 
ratio (SNR) and lead to the considerable pilot overhead. In order to reduce the high pilot overhead in channel estimation, in the current massive 
MIMO system, some algorithms based on compressed sensing, such as orthogonal matching pursuit (OMP) \cite{b4} and sparse Bayesian 
learning \cite{b5}, are expliored by taking advantage of channel sparsity in the angle domain to estimate high-dimensional channels 
with low pilot overhead. However, these methods require prior knowledge and often have higher complexity when dealing with 
high-dimensional channel matrices.

In recent years, with the rapid advancement of deep learning (DL) technology, researchers have begun exploring its potential 
applications in channel estimation. DL models can automatically learn complex feature representations, significantly 
improving the accuracy and robustness of channel estimation \cite{b15} --\cite{b17}. Specifically, several prior works have employed convolutional neural networks 
(CNNs) for channel estimation \cite{b6}. In \cite{b9}, recurrent neural networks (RNNs) were utilized to process historical channel inputs iteratively sequentially.

The attention mechanism\cite{b10}, as a milestone method of DL, has been widely applied in fields such as natural language 
processing \cite{b11} and computer vision. The attention mechanism 
dynamically assigns different weights to different parts of the input 
data, enabling the model to focus on more relevant information, which is crucial for 
enhancing the model's performance and efficiency.

The main contributions of this paper can be summarized as follows:

1) Faced with the complexity of high-dimensional channel estimation in XL-MIMO systems and the challenges of hybrid-field, we 
proposed a
mixed attention transformer based channel estimation neural
network (MAT-CENet). Leveraging the model architecture and insightful training data generation 
method, MAT-CENet is applicable to hybrid-field users with different channel statistics, without relying on prior knowledge.

2) The proposed structure combines CNNs and transformer encoder structure, enhancing feature extraction and representation capabilities. This integration enables the model to capture complex relationships,
significantly improving the accuracy and stability of channel estimation.

3) Extensive simulations were conducted along with comparative analyses to other methods to demonstrate the superior performance of 
MAT-CENet in estimating the XL-MIMO channel.

\begin{figure}[t]
\centerline{\includegraphics[scale=0.5]{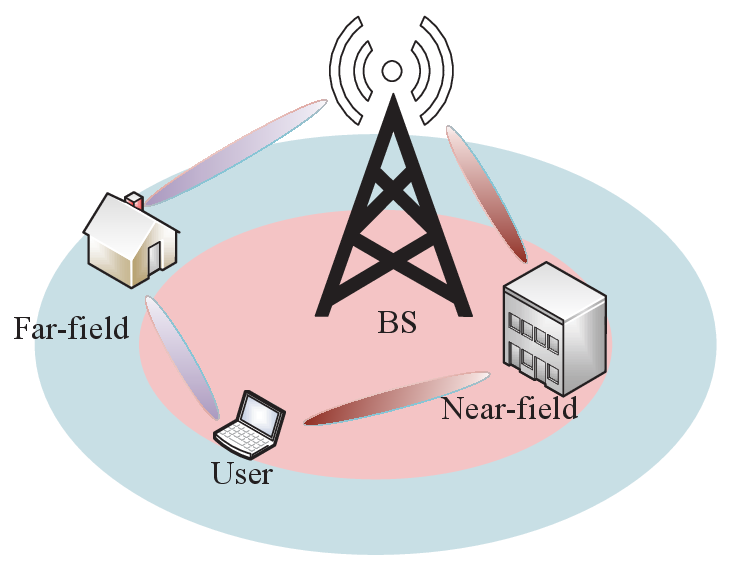}}
\caption{A hybrid-field XL-MIMO system. }
\label{figure1}
\end{figure}

\section{System model}

As shown in Fig.~\ref{figure1}, consider that the base station (BS) of an XL-MIMO system equipped with \(M \) antennas serves a user 
with a single antenna, with \(L\)  paths between the base station and the user.

We denote \(\mathbf{h}  \in \mathbb{C}^{M\times1 }\) as the channel vector between the base station and the user. 
\(x \in \mathbb{C}\) represents the pilot signal transmitted by the user within a time slot, and \(\mathbf{\boldsymbol{\eta} } \in \mathbb{C}^{M\times1}\)
represents the additive Gaussian noise. Thus, the uplink received
signals at the BS \(\mathbf{y}\) can
be formulated by

\begin{equation}
  \mathbf{y = h} x \mathbf{+ \boldsymbol{\eta}}. \label{eq1}
\end{equation}

Prior research on efficient channel estimation in XL-MIMO systems generally considers the channel models as either far-field or near-field.
In the XL-MIMO system, the separation between near-field and far-field propagation effects is determined by the Rayleigh distance, given by
\(
D_{{Ray}} = \frac{2D_{{a}}^2}{\lambda}\label{eq3}\), where \( D_a \)  is the array aperture and \( \lambda \) is the carrier wavelength. 
For a uniform linear array,  
if the spacing between adjacent antennas is \( d = \frac{\lambda}{2} \), the Rayleigh distance \( D_{Ray} \) can be expressed as

\begin{equation}
D_{{Ray}} = \frac{M^2 \lambda}{2}\label{eq2}.
\end{equation}
The radius of \( D_{Ray} \) is sufficient to cover a portion of users, 
causing them to experience the near-field propagation effect. 

As depicted in Fig.~\ref{figure1}, when the scattering distance is larger than \( D_{Ray} \), the channel is modeled using 
the far-field formula expressed as
\begin{equation}
  \mathbf{h}_f = \sqrt{\frac{M}{L}}  \sum_{l=1}^{L} g_{l} \mathbf{a}(\phi_l), \label{eq3}
\end{equation}
where \( \mathbf{a}(\phi_l) = \frac{1}{\sqrt{M}}[1, e^{-j2\pi \frac{d}{\lambda} \sin \phi_l},\ldots, e^{-j2\pi \frac{d}{\lambda} (M-1) \sin \phi_l}]\)
is the array response vector of the base station antenna array for the \( l \)-th path, 
and \( \phi_l \in [-\frac{\pi}{2}, \frac{\pi}{2}] \) and \(g_{l}\) denote the azimuth angle and the gain of the \( l \)-th path, respectively.

When the scattering distance is smaller than \( D_{Ray} \), the channel is modeled using 
the near-field formula
\begin{equation}
  \mathbf{h}_n = \sqrt{\frac{M}{L}}  \sum_{l=1}^{L} g_{l} 
  \mathbf{a}(\phi_l, r_l), \label{eq4}
\end{equation}
where \( \mathbf{a}(\phi_l, r_l) = \frac{1}{\sqrt{M}}[1, e^{-j2\pi \frac{d}{\lambda} (r_l,1-r_l))},\ldots,e^{-j2\pi \frac{d}{\lambda} (r_l,M-r_l)}]\) is the antenna array response vector with the azimuth angle \( \phi_l \) and the 
distance \( r_l \) for the \( l \)-th path. The distance \( r_l \) represents the distance from the \( l \)-th scatterer 
to the center of the base station antenna array , \(
r_{l,m} = \sqrt{r_l^2 + \delta_m^2 d^2 - 2 r_l \delta_m d \sin \phi_l}
\) and \( \delta_m = 2m - \frac{M - 1}{2} \) represents the relative position of the \( m \) th antenna with respect to 
the center of the array.

However, in some application scenarios, the scatters may contain both near-field and far-field components. In such cases, 
the propagation and reception of the signal will be influenced by the hybrid 
effects of both field regions. The hybrid-field channel model can be expressed as
\begin{equation}
\begin{split}
\mathbf{h }&= \mathbf{h}_f + \mathbf{h}_n \\
 &= \sqrt{\frac{M}{L}} \left[ \sum_{l=1}^{L_0} g_{l} \mathbf{a}(\phi_l) + \sum_{l=L_0+1}^{L} g_{l} 
\mathbf{a}(\phi_l, r_l) \right].\label{equ5}
\end{split}
\end{equation}
The above equation shows that the hybrid-field contains \( L_0 \) far-field paths and \( L - L_0 \) near-field paths.

\section{Mixed-Attention Aided DL Framework for Channel Estimation}
In this section, we first introduce the preprocessing procedure. Next, we provide a explanation of MAT-CENet framework, 
followed by  a description of the Transformer encoder structure. Finally, we present the computational complexity analysis.
\subsection{Preprocessing}

\begin{figure*}[htbp]
  \centerline{\includegraphics[width=\textwidth,height=10cm]{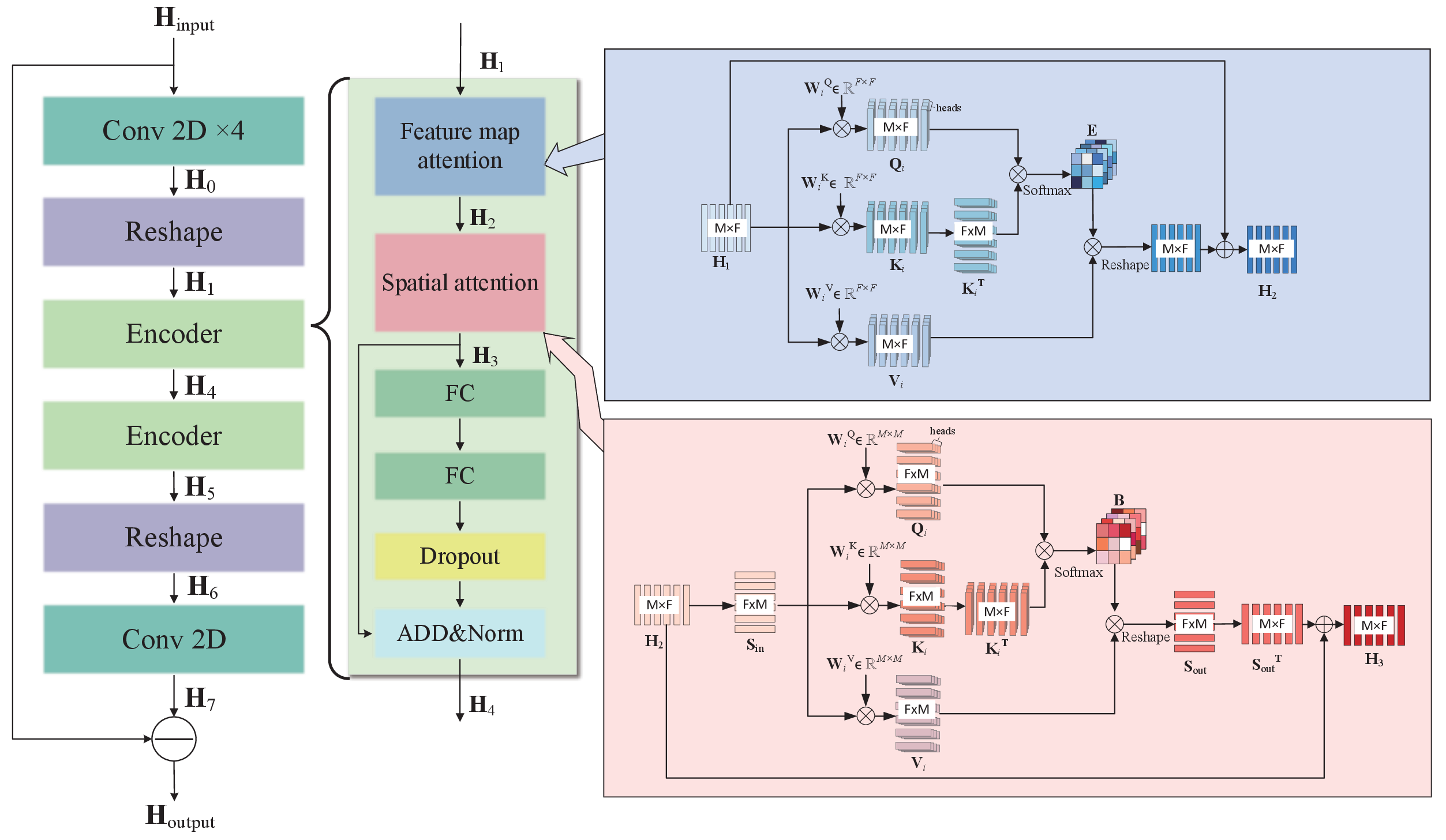}}
  \caption{ Overall structure of MAT-CENet. }
  \label{fig2}
\end{figure*}

Given that \( x \) in (\ref{eq1}) is a pilot signal with a known power, and assuming its power is \( P \), we can express 
\( x\) as \( x = \sqrt{P} \tilde{x} \), where \( \tilde{x} \) is a pilot signal with unit power. Thus, without loss of generality, 
we can simplify (\ref{eq1}) to
\begin{equation}
  \mathbf{y }= \sqrt{P} \mathbf{h + \boldsymbol{\eta}}. \label{eq6}
\end{equation} 
Based on the above equation, the LS estimate of \(\mathbf{h}\) is expressed by
\begin{equation}
  \mathbf{\hat{h}} = \frac{\mathbf{y}}{\sqrt{P}} = \mathbf{h} + \frac{\mathbf{\boldsymbol{\eta}}}{\sqrt{P}}. \label{eq7}
\end{equation}
The complex matrix \( \mathbf{\hat{h}} \) is split into real and imaginary parts of the same dimension, then concatenated and 
reshaped into a real-valued matrix \( \mathbf{H}_\text{input} \) with dimensions \( (\sqrt{M}, \sqrt{M },2) \) as the input to 
MAT-CENet, where the square root of \(M\) is assumed existent without loss of generality.

\subsection{Overall Structure of the Proposed Network}
As shown in Fig.~\ref{fig2}, the input tensor \( \mathbf{H}_\text{input} \in \mathbb{R}^{\sqrt{M} \times\sqrt{M} \times 2}\) is first fed 
sequentially through four convolutional blocks to generate the feature map 
\( \mathbf{H}_0 \in \mathbb{R}^{\sqrt{M} \times\sqrt{M} \times F}\), where \(F\) represents the number of feature maps in the tensor. 
Each convolutional block consists of a two-dimensional (2D) convolutional layer with 64 filters of size \( 3 \times 3 \), batch normalization (BN), 
and  rectified linear unit (ReLU) activation functions, which are used for feature extraction and improving 
training stability and convergence speed in channel estimation. \( \mathbf{H}_0\) is then  reshaped into a vector 
\( \mathbf{H}_1 \in \mathbb{R}^ {M \times F}\) suitable for the encoder input.

Next, \( \mathbf{H}_1 \) passes through two encoder structures sequentially to obtain the tensor \( \mathbf{H}_5 \in \mathbb{R}^ {M \times F}\), 
capturing relationships between different features and spatial positions. After that, \( \mathbf{H}_5 \) is reshaped into 
\( \mathbf{H}_6  \in \mathbb{R}^{\sqrt{M} \times\sqrt{M} \times F}\) to be processed by the subsequent 2D convolutional layer with 
2 filters of size 3×3 and BN, generating \( \mathbf{H}_7  \in \mathbb{R}^{\sqrt{M} \times\sqrt{M} \times 2}\). Finally, the input tensor 
\( \mathbf{H}_\text{input} \)  and the output tensor \( \mathbf{H}_7\) from the convolutional layer are passed through a residual 
network layer, which accelerates the network’s convergence and enhances its expressive capability, yielding in the final output 
\( \mathbf{H}_\text{output} \).

The proposed model integrates  several convolutional layers with mixed-domain attention layers to achieve the improved feature extraction. 
This architecture is optimized to maintain a good balance between accuracy and computational complexity. In contrast, other 
mixed-domain attention mechanism approaches may differ in model structure, such as using alternative convolutional layer 
configurations or attention mechanism implementations. These models might not be specifically optimized for XL-MIMO 
channels, leading to the unsatisfactory performance.

\subsection{Encoder Structure of MAT-CENet }
The encoder structure designed in this paper utilizes multi-head attention applied to both feature dimensions and 
spatial position dimensions to enhance the model ability to capture comprehensive characteristics of the input data. 
After passing through the two attention modules, the output \( \mathbf{H}_3 \in \mathbb{R}^ {M \times F}\) is processed by a feed forward 
neural network composed of two fully connected layers. The first fully connected (FC) layer introduces non-linearity through the ReLU 
activation function, enabling the model to learn more complex representations. The second FC layer has an output dimension 
matching that of \( \mathbf{H}_3\), ensuring consistency in the dimension for 
subsequent residual connections. Finally, layer normalization is applied to ensure the stability of the output 
\( \mathbf{H}_4\in \mathbb{R}^ {M \times F}\).

In channel estimation, the feature map attention mechanism and the spatial attention mechanism serve distinct roles and focus on different aspects. They work together to enhance the model's ability to capture and estimate channel characteristics.

\subsubsection{Feature Map Attention}
By leveraging the interdependencies between feature maps, we can enhance the crucial features for channel estimation while ignoring 
noise or secondary features. Therefore, we have constructed a feature map attention module, the structure of which is shown in Fig.~\ref{fig2}. 
Specifically, we use the 
input matrix \( \mathbf{H}_1 \) and the multi-head attention mechanism to generate \( h \) different feature weight matrices 
\( \mathbf{E}\in \mathbb{R}^ {F \times F} \), where \( h \) denotes the number of heads in multi-head attention mechanism. 
These matrices are then multiplied with the \( h \) matrices obtained by processing \( \mathbf{H}_1 \) 
through the attention mechanism, and the results are concatenated and linearly transformed to produce a matrix with the same dimensions 
as \( \mathbf{H}_1 \). Finally, an element-wise summation  between the resulting reshaped matrix and \(\mathbf{H}_1 \) 
is performed to obtain the 
final output \( \mathbf{H}_2 \in \mathbb{R}^ {M \times F} \). \( \mathbf{H}_2\) represents the weighted sum of all these high-dimensional
feature maps. The feature map attention module enhances the 
model's focus on important features by assigning weights to the features and suppressing secondary ones, thereby improving the accuracy 
of channel estimation.

\subsubsection{Spatial Attention}
The spatial position attention mechanism focuses more on the distribution and relationships of features in the spatial dimension 
by modeling the relationships between different spatial positions to capture the spatial variation patterns of channel features. 
Unlike the feature map attention mechanism, the spatial 
attention module shown in Fig.~\ref{fig2} cannot directly use the input matrix. 
To extract the spatial feature weight matrix \( \mathbf{B}\in \mathbb{R}^ {M \times M}\), the input 
matrix \( \mathbf{H}_2\) needs to be transposed to swap dimensions, allowing \( \mathbf{B}\) to act on the spatial dimension with the 
transposed matrix \( \mathbf{S}_\text{in} \in \mathbb{R}^ {F \times M}\). Finally, the attention output matrix 
\( \mathbf{S}_\text{out} \in \mathbb{R}^ {F \times M}\) is transposed again and element-wise summed with \( \mathbf{H}_2\) to obtain 
the output \( \mathbf{H}_3 \in \mathbb{R}^ {M \times F}\). \( \mathbf{H}_3\) is the weighted sum of all position features and the 
original features at each position.
The spatial position attention module helps the model recognize important spatial correlations, 
thereby improving the accuracy of channel estimation.

The attention mechanism is a fundamental concept in deep learning, which enables the model to dynamically focus on different parts of 
the input data, allowing it to effectively handle long-range dependencies and complex relationships. 
To be specific, use \(\mathbf{X} = [x_1, x_2, \cdots, x_n ]\) to represent \(n\) input information. Through linear transformations, 
obtain the initial representations of the vectors queries (\( \mathbf{Q}\)), keys (\( \mathbf{K} \)), and values (\( \mathbf{V} \)), 
with \( \mathbf{W} \) representing the corresponding weight matrices.
\(\mathbf{Q = W}_\text{q}  \mathbf{X}  \),  \(
\mathbf{K = W}_\text{k} \mathbf{X}  
\), \(
\mathbf{V = W}_\text{v} \mathbf{X}  
\).
The formula for scaled dot-product attention is
\begin{equation}
{Attention}(\mathbf{Q, K, V}) = \text{softmax} \left( \frac{\mathbf{QK}^T}{\sqrt{d_k}} \right) \mathbf{V}, \label{eq8}
\end{equation}
where \( d_k \) is the dimension of \(\mathbf{ K}\).

In the feature map attention module and the spatial attention module depicted in Fig.~\ref{fig2}, we employ the multi-head attention 
mechanism. Instead of performing a single attention function, multi-head attention runs 
several attention mechanisms in parallel, called heads. \(\mathbf{Q}_i = \mathbf{QW}_i^\text{Q}, \quad \mathbf{K}_i = 
\mathbf{KW}_i^\text{K}, \quad \mathbf{V}_i = \mathbf{VW}_i^\text{V}\) represent the generation of  subspace representations for the 
Query, Key, and Value through their respective linear transformation matrices \(\mathbf{W}_i\).
The formula for multi-head attention is
\begin{equation}
\mathbf{head}_i = {Attention}(\mathbf{Q}_i, \mathbf{K}_i, \mathbf{V}_i), \quad i = 1, \ldots, h \label{eq9}
\end{equation}
\begin{equation}
{Multi}(\mathbf{Q}, \mathbf{K}, \mathbf{V}) = {Concat}(\mathbf{head}_1, \ldots, \mathbf{head}_h) \mathbf{W}^O. \label{eq10}
\end{equation} 
Each attention head processes the input data independently and performs scaled dot-product attention. Their outputs are 
concatenated to form a new output, which is then linearly transformed to produce the final multi-head attention output.
The multi-head attention enables the model to capture different subspace features of the input data by processing multiple 
attention heads in parallel, thus better modeling complex dependencies. Each head focuses on different parts or features, 
making multi-head attention more powerful and flexible than single-head attention.

Compared to other transformer encoder structures, the encoder structure of MAT-CENet leverages multi-head attention mechanisms to enhance the model's 
feature extraction capabilities. By combining feature map and spatial attention mechanisms, the MAT-CENet encoder can process information 
in parallel across two different dimensions, making the model more robust in capturing and expressing fine-grained features of the data.

\subsection{Complexity Analysis}

In this subsection, the computational complexity of the proposed model during the inference stage is analyzed. 
The complexity is evaluated in terms of floating-point operations (FLOPs). There are two typical network
structures that require computational FLOPs in the proposed framework. \par
\subsubsection{Convolutional Layers}
The first four convolutional layers are the initial layers that the input \(\mathbf{H}_{\text{input}}\) passes 
through, while the fifth convolutional layer is the one before the subtraction operation of the residual structure.
The computational complexity of these five convolutional layers is  
\begin{equation}
\mathcal{C}_{\text{convs}} \sim \mathcal{O}(\sum_{i=1}^{5}N_x N_y K^2 C_{i\text{,in}} C_{i\text{,out}}),\label{eq11}
\end{equation}
where \(N_x \) and \(N_y \) denote
the length and width of output feature maps, \(K \) denotes the side length of the kernel, \(C_{i\text{,in}} \) and \(C_{i\text{,out}} \) 
denote the numbers of input and output feature maps in the \(i\)-th convolutional layer.

\subsubsection{Encoders} Another is the encoder module whose computational complexity is mainly affected by multi-attention layers and FC layers.
The computational complexity of multi-attention layer is
\begin{equation}
\mathcal{C}_{\text{att}} \sim \mathcal{O}(D_{\text{model}}^2 D_{\text{head}} h + D_{\text{model}} D_{\text{head}} h ),\label{eq12}
\end{equation}
where \(D_{\text{model}}\), \(D_{\text{head}}\) and  \(h\) denote the dimension of the input sequence, the dimension of each 
attention head and the number of attention heads is \(h\). Since in this paper, \(D_{\text{model}}\) = \(D_{\text{head}} \) = \(D_{\text{feature}}\) 
in the channel attention, and \(D_{\text{model}}\) = \(D_{\text{head}} \) = \(D_{\text{spatial}}\) in the spatial attention.
The computational complexity of FC layer is
\begin{equation}
\mathcal{C}_{\text{FC}} \sim \mathcal{O}( Z_{\text{in}} Z_{\text{out}}),\label{eq13}
\end{equation}
where \( Z_{\text{in}}\) and \( Z_{\text{out}}\) denote the numbers of input and output neurons, respectively.

Combining the multi-attention layers and FC layers, the computational complexity of encoder is
\begin{equation}
\mathcal{C}_{\text{encoder}} \sim 2 \times \mathcal{C}_{\text{att}}+2 \times \mathcal{C}_{\text{FC}}.\label{eq14}
\end{equation}

The overall construction contains 5 convolutional layers and 2 encoder modules. 
\begin{equation}
  \begin{split}
    \mathcal{C}_{\text{MAT-CENet}} \sim \mathcal{C}_{\text{convs}}+2 \times \mathcal{C}_{\text{encoder}}\label{eq15}
  \end{split}
  \end{equation}

According to (\ref{eq11}) -- (\ref{eq15}), the
overall complexity is generally expressed by 
\begin{equation}
\begin{split}
  \mathcal{C}_{\text{MAT-CENet}} \sim  \mathcal{O}(&\sum_{i=1}^{5}N_x N_y K^2 C_{i\text{,in}} C_{i\text{,out}}+2D_{\text{feature }}^3 h\\
  &+ 2D_{\text{feature}} ^2 h + 2D_{\text{spatial }}^3 h+ 2D_{\text{spatial}} ^2 h\\
  & + 4F_{\text{in}} F_{\text{out}}).\label{eq16}
\end{split}
\end{equation}
Thus, the computational complexity of the proposed model is primarily determined by the dimensions of the input matrices 
\(N_x\) and \(N_y\), and the dimension of each attention head.

Here is the parameter count and FLOPs of the proposed model 
structure and the structure XLCNet \cite{b13} used for simulation comparison in  shown in Table 1. XLCNet is an XL-MIMO channel network composed of 9 convolutional layers and a residual structure, where 8 of the convolutional layers have 64 filters of size 3×3 each, and the remaining one has 2 filters of size 3×3.
\begin{table}[htbp]
\caption{complexity comparison}
\begin{center}
\begin{tabular}{c|c|c} 
\hline
\textbf{ } & \textbf{\textit{Parameters}} & \textbf{\textit{FLOPs}} \\
\hline
{XLCNet} & {263K} & {67M}\\
\hline
{MAT-CENet} & {2.38M} & {165M}\\
\hline
\end{tabular}
\label{tab1}
\end{center}
\end{table}

Although MAT-CENet has higher complexity compared to XLCNet, subsequent simulations show that it provides higher prediction accuracy.

\section{Simulation Result}
In this section, we first introduce the preparations and data processing done for the simulation. The second part presents 
the simulation results and analyzes the results by comparing MAT-CENet with other schemes.
\subsection{Simulation Setup}
Before training and testing with the proposed network, we first generate the required training and testing datasets based on 
equations (1) and (5). Choosing appropriate values for \( L_0 \) and \( L \) can help improve MAT-CENet training efficiency and generalization ability 
in scenarios with different propagation effects and path numbers. We then train MAT-CENet
to minimize the mean squared error (MSE) loss.
\begin{equation}
\mathbf{\mathcal{L}} = \frac{1}{N_{\text{tr}}} \sum_{n=1}^{N_{\text{tr}}} \left\| \mathbf{\bar{h}^{(n)} - \hat{\bar{h}}^{(n)} }\right\|_F^2
  \label{eq17}
\end{equation}
In (\ref{eq17}), \( N_{\text{tr}} \) represents the number of training samples, and the superscript \( (n) \) indicates the 
\( n \)th sample. As a performance indicator, the normalized mean square error (NMSE) defined as \(\mathbf{NMSE} =\mathbb{E}
\{\frac{\| \bar{h} - \hat{\bar{h}} \|_F^2}{\| \bar{h} \|_F^2}
\}
\)
estimation performance.

The simulation parameters are configured as follows: the number of BS antennas \(M \) = 256, the wavelength \( \lambda = 0.01\)m, 
the mean path gain \(\sigma^2 = 1\), \(\quad \phi_l \sim U \left(-\frac{\pi}{2}, \frac{\pi}{2}\right)\), 
\(\text{ and } r_l \sim U(10, 80)\) meters. In the simulation, the training set has 9000 samples, and the verification 
set has 1000 samples, and both are formed by a hybrid field channel model. 
The test sets for Fig.~\ref{figure3} and Fig.~\ref{figure4} are formed when there are only near-field path channels or far-field path channels. The total number of hybrid-field paths is set to 6, and the number of far-field paths is increased to form the test sets for Fig.~\ref{figure5}. Adam is applied as the 
optimizer, and the batch size is set to 128. The number
of training epochs and the learning rate are set to \(N_{epoch}\)= 200
and  \(
\alpha  = 10^{-3}
\), respectively.  
Specifically, we evaluate the proposed method against the conventional LS estimation method, the linear minimum mean-squared error (LMMSE) estimation method,  hybrid-field orthogonal matching pursuit (HY-OMP) \cite{b14} 
and an XL-MIMO channel network (XLCNet).

\subsection{Simulation Result}
The following discussions  provide a detailed comparison of channel estimation performance between the MAT-CENet network and other schemes  
under various SNR conditions. The horizontal axis represents the SNR values, ranging from -10 dB to 20 dB. 
“MAT-CENet” and “XLCNet” are trained using the same dataset with 1 far-field path and 5 near-field
paths.

\begin{figure}[t]
  \centerline{\includegraphics[scale=0.75]{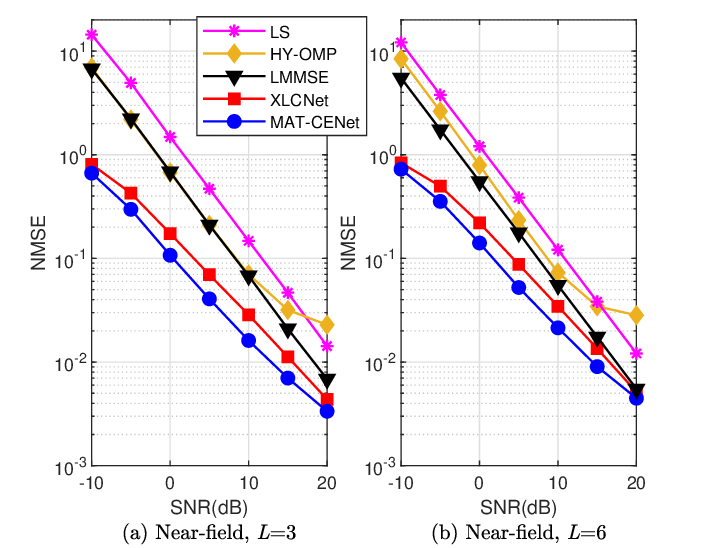}}
  \caption{NMSE versus SNR for the near-field user. }
  \label{figure3}
  \end{figure}
  \begin{figure}[t]
  \centerline{\includegraphics[scale=0.75]{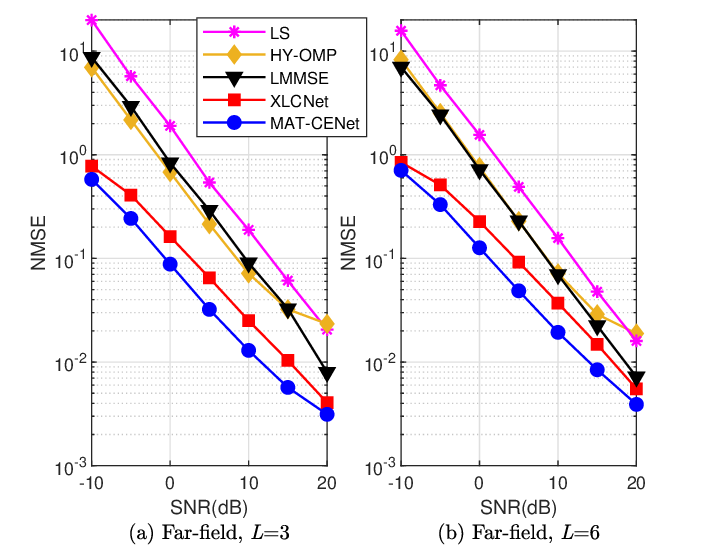}}
  \caption{NMSE versus SNR for the far-field user.  }
  \label{figure4}
  \end{figure}

  \begin{figure}[t]
    \centerline{\includegraphics[scale=0.75]{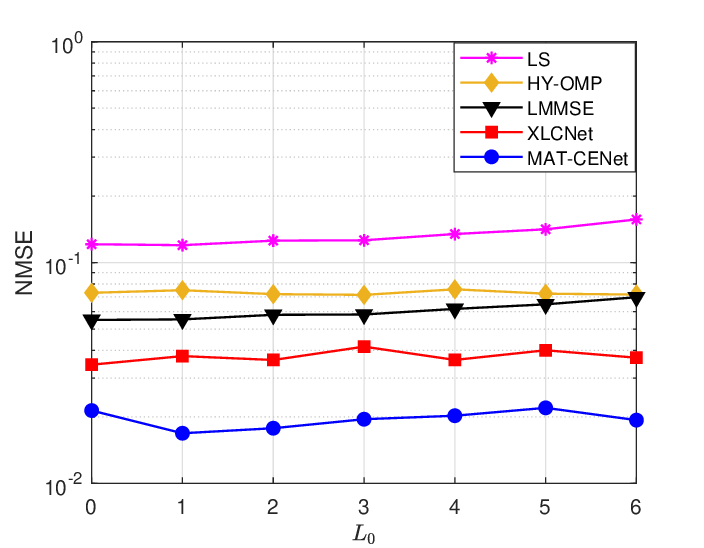}}
    \caption{NMSE versus  the number of far-field paths \( L_0 \) for the hybrid-field with \( L \) = 6. }
    \label{figure5}
  \end{figure}

Fig.~\ref{figure3} depicts the NMSE performance versus SNR for the near-field user, 
where the two subfigures correspond to 3 and 6 near-field paths. The MAT-CENet exhibits outstanding performance under near-field path conditions. 
It significantly reduces estimation errors, especially at low SNR levels, demonstrating higher sensitivity and accuracy 
in near-field signal processing. Despite the increase in path numbers in Fig.~\ref{figure3} (b) , the MAT-CENet maintains high 
estimation accuracy. Fig.~\ref{figure4} shows the NMSE performance versus SNR for the far-field user, 
where the two subfigures correspond to 3 and 6 far-field paths.  In Fig.~\ref{figure4} (a) and (b), the MAT-CENet  
consistently outperforms other schemes across most SNR levels, indicating 
its effectiveness in extracting channel features in multi far-field path scenarios. Although the overall estimation error 
increases with the number of paths, the MAT-CENet retains its advantage, especially at higher SNR levels.

Fig.~\ref{figure5} illustrates the performance of various schemes in terms of NMSE as the number of far-field paths 
increases, with SNR = 10dB and a total of 6 paths. As depicted in the figure, MAT-CENet consistently outperforms other schemes in 
hybrid-field scenarios, regardless of the number of far-field paths. Furthermore, since the training dataset was generated using a 
simulation channel model with 1 far-field path and 5 near-field paths, the model achieves optimal performance in this specific channel 
estimation task. Nevertheless, MAT-CENet also demonstrates robust performance in other hybrid-field channel estimations, highlighting the 
model's strong generalization capability.

In all comparison schemes, it can be seen that the proposed network architecture MAT-CENet outperforms other baseline schemes in the 
vast majority of cases. MAT-CENet and XLCNet do not rely on channel statistical knowledge , which is an advantage for channel estimation. 
The performance advantage 
of MAT-CENet over XLCNet reveals that network architectures with attention mechanisms are more superior for channel estimation tasks.

\section{Conclusion}
In this paper, we propose a channel estimation neural network based on the mixed attention transformer, aiming to eliminate the 
reliance on channel prior knowledge and to capture the unique characteristics of XL-MIMO channels more effectively. Compared to 
other channel estimation schemes, MAT-CENet outperforms comparative methods in near-field, far-field and hybrid-field propagation 
scenarios, demonstrating the exceptional performance and generalization capability.

\vspace{12pt}

\end{document}